\documentclass[aip,jap,preprint,showpacs,superscriptaddress,groupedaddress]{revtex4-1}  
\usepackage{graphicx}  
\usepackage{bm}        
\usepackage{amssymb}   
\usepackage{times}
\usepackage[version=3]{mhchem}
\usepackage{lineno}

\begin{document}

\title{Interfacial Thermal Conductance of Thiolate-Protected
  Gold Nanospheres}
\author{Kelsey M. Stocker}
\author{Suzanne M. Neidhart}
\author{J. Daniel Gezelter}
\email{gezelter@nd.edu}
\affiliation{Department of Chemistry and Biochemistry, University of
  Notre Dame, Notre Dame, IN 46556}

\begin{abstract}
  Molecular dynamics simulations of thiolate-protected and solvated
  gold nanoparticles were carried out in the presence of a
  non-equilibrium heat flux between the solvent and the core of the
  particle. The interfacial thermal conductance ($G$) was computed for
  these interfaces, and the behavior of the thermal conductance was
  studied as a function of particle size, ligand flexibility, and
  ligand chain length. In all cases, thermal conductance of the
  ligand-protected particles was higher than the bare metal--solvent
  interface.  A number of mechanisms for the enhanced conductance were
  investigated, including thiolate-driven corrugation of the metal
  surface, solvent ordering at the interface, solvent-ligand
  interpenetration, and ligand ordering relative to the particle
  surface. Only the smallest particles exhibited significant
  corrugation.  All ligands permitted substantial solvent-ligand
  interpenetration, and ligand chain length has a significant
  influence on the orientational ordering of interfacial solvent.
  Solvent -- ligand vibrational overlap, particularly in the low
  frequency range ($< 80 \mathrm{cm}^{-1}$) was significantly altered
  by ligand rigidity, and had direct influence on the interfacial
  thermal conductance.
\end{abstract}

\maketitle

\section{Introduction}

Heat transport across various nanostructured interfaces has been the
subject of intense experimental
interest,\cite{Wilson:2002uq,Ge:2004yg,Shenogina:2009ix,Wang10082007,Schmidt:2008ad,Juve:2009pt,Alper:2010pd,Harikrishna:2013ys}
and the interfacial thermal conductance, $G$, is the principal
quantity of interest for understanding interfacial heat
transport.\cite{Cahill:2003fk} Because nanoparticles have a
significant fraction of their atoms at the particle / solvent
interface, the chemical details of these interfaces govern the thermal
transport properties.  Time-domain thermoreflectance (TDTR)
measurements on planar self-assembled monolayer (SAM) junctions
between quartz and gold films showed that surface chemistry,
particularly the density of covalent bonds to the gold surface, can
control energy transport between the two solids.\cite{Losego:2012fr}
Experiments and simulations on three-dimensional nanocrystal arrays
have similarly shown that surface-attached ligands mediate the thermal
transport in these materials, placing particular importance on the
overlap between the ligand and nanoparticle vibrational densities of
states.\cite{Ong:2013rt,Ong:2014yq} Likewise, simulations of
polymer-coated gold nanoparticles in water have shown that the surface
coating introduces a dominant thermal transport channel to the
surrounding solvent.\cite{Soussi:2015fj}

For ligand-protected nanoparticles in a solvent, there may be three
distinct heat transfer processes: (1) from the particles to the
ligands, (2) vibrational energy tranfer along the length of the
ligand, followed by (3) heat transport from the ligand to the
surrounding solvent.\cite{Ge:2006kx}

Heat transport at the gold-alkylthiolate-solvent interface has been
previously explored both through molecular dynamics simulations and
via
TDTR.\cite{Kikugawa:2009vn,Kuang:2011ef,Stocker:2013cl,Tian:2015uq}
Most of these studies have found that alkylthiolates enhance the
thermal conductance to the solvent, and that the vibrational overlap
provided by the chemically-bound ligand species plays a role in this
enhancement.

Reverse nonequilibrium molecular dynamics (RNEMD)
methods~\cite{Muller-Plathe:1997wq} have been previously applied to
calculate the thermal conductance at flat (111) metal / organic
solvent interfaces that had been chemically protected by varying
coverages of alkanethiolate groups.\cite{Kuang:2011ef} These
simulations suggested an explanation for the increased thermal
conductivity at alkanethiol-capped metal surfaces compared with bare
metal interfaces.  Specifically, the chemical bond between the metal
and the ligand introduces a vibrational overlap that is not present
without the protecting group, and the overlap between the vibrational
spectra (metal to ligand, ligand to solvent) provides a mechanism for
rapid thermal transport across the interface. The simulations also
suggested that this phenomenon is a non-monotonic function of the
fractional coverage of the surface, as moderate coverages allow energy
transfer to solvent molecules that come into close contact with the
ligands.

Similarly, simulations of {\it mixed-chain} alkylthiolate surfaces
showed that solvent trapped close to the interface can be efficient at
moving thermal energy away from the surface.\cite{Stocker:2013cl}
Trapped solvent molecules that were orientationally aligned with
nearby ligands were able to increase the thermal conductance of the
interface.  This indicates that the ligand-to-solvent vibrational
energy transfer is a key feature for increasing particle-to-solvent
thermal conductance.

Recently, we extended RNEMD methods for use in non-periodic geometries
by creating scaling/shearing moves between concentric regions of a
simulation.\cite{Stocker:2014qq} In this work, we apply this
non-periodic variant of RNEMD to investigate the role that {\it
  curved} nanoparticle surfaces play in heat and mass transport.  On
planar surfaces, we discovered that orientational ordering of surface
protecting ligands had a large effect on the heat conduction from the
metal to the solvent.  Smaller nanoparticles have high surface
curvature that creates gaps in well-ordered self-assembled monolayers,
and the effect of those gaps on the thermal conductance is unknown.


For a solvated nanoparticle, it is possible to define a critical value
for the interfacial thermal conductance,
\begin{equation}
G_c = \frac{3 C_s \Lambda_s}{R C_p}
\end{equation}
which depends on the solvent heat capacity, $C_s$, solvent thermal
conductivity, $\Lambda_s$, particle radius, $R$, and nanoparticle heat
capacity, $C_p$.\cite{Wilson:2002uq} In the limit of infinite
interfacial thermal conductance, $G \gg G_c$, cooling of the
nanoparticle is limited by the solvent properties, $C_s$ and
$\Lambda_s$.  In the opposite limit, $G \ll G_c$, the heat dissipation
is controlled by the thermal conductance of the particle / fluid
interface. It is this regime with which we are concerned, where
properties of ligands and the particle surface may be tuned to
manipulate the rate of cooling for solvated nanoparticles.  Based on
estimates of $G$ from previous simulations as well as experimental
results for solvated nanostructures, gold nanoparticles solvated in
hexane are in the $G \ll G_c$ regime for radii smaller than 40 nm. The
particles included in this study are more than an order of magnitude
smaller than this critical radius, so the heat dissipation should be
controlled entirely by the surface features of the particle / ligand /
solvent interface.

\subsection{Structures of Self-Assembled Monolayers on Nanoparticles}

Though the ligand packing on planar surfaces has been characterized
for many different ligands and surface facets, it is not obvious
\emph{a priori} how the same ligands will behave on the highly curved
surfaces of spherical nanoparticles. Thus, as new applications of
ligand-stabilized nanostructures have been proposed, the structure and
dynamics of ligands on metallic nanoparticles have been studied using
molecular simulation,\cite{Henz:2008qf} NMR, XPS, FTIR,
calorimetry, and surface
microscopies.\cite{Badia1996:2,Badia1996,Badia1997:2,Badia1997,Badia2000}
Badia, \textit{et al.} used transmission electron microscopy to
determine that alkanethiol ligands on gold nanoparticles pack
approximately 30\% more densely than on planar Au(111)
surfaces.\cite{Badia1996:2} Subsequent experiments demonstrated that
even at full coverages, surface curvature creates voids between linear
ligand chains that can be filled via interdigitation of ligands on
neighboring particles.\cite{Badia1996} The molecular dynamics
simulations of Henz, \textit{et al.} indicate that at low coverages,
the thiolate alkane chains will lie flat on the nanoparticle
surface\cite{Henz:2008qf} Above 90\% coverage, the ligands
stand upright and recover the rigidity and tilt angle displayed on
planar facets. Their simulations also indicate a high degree of mixing
between the thiolate sulfur atoms and surface gold atoms at high
coverages.

\begin{figure}
  \includegraphics[width=\linewidth]{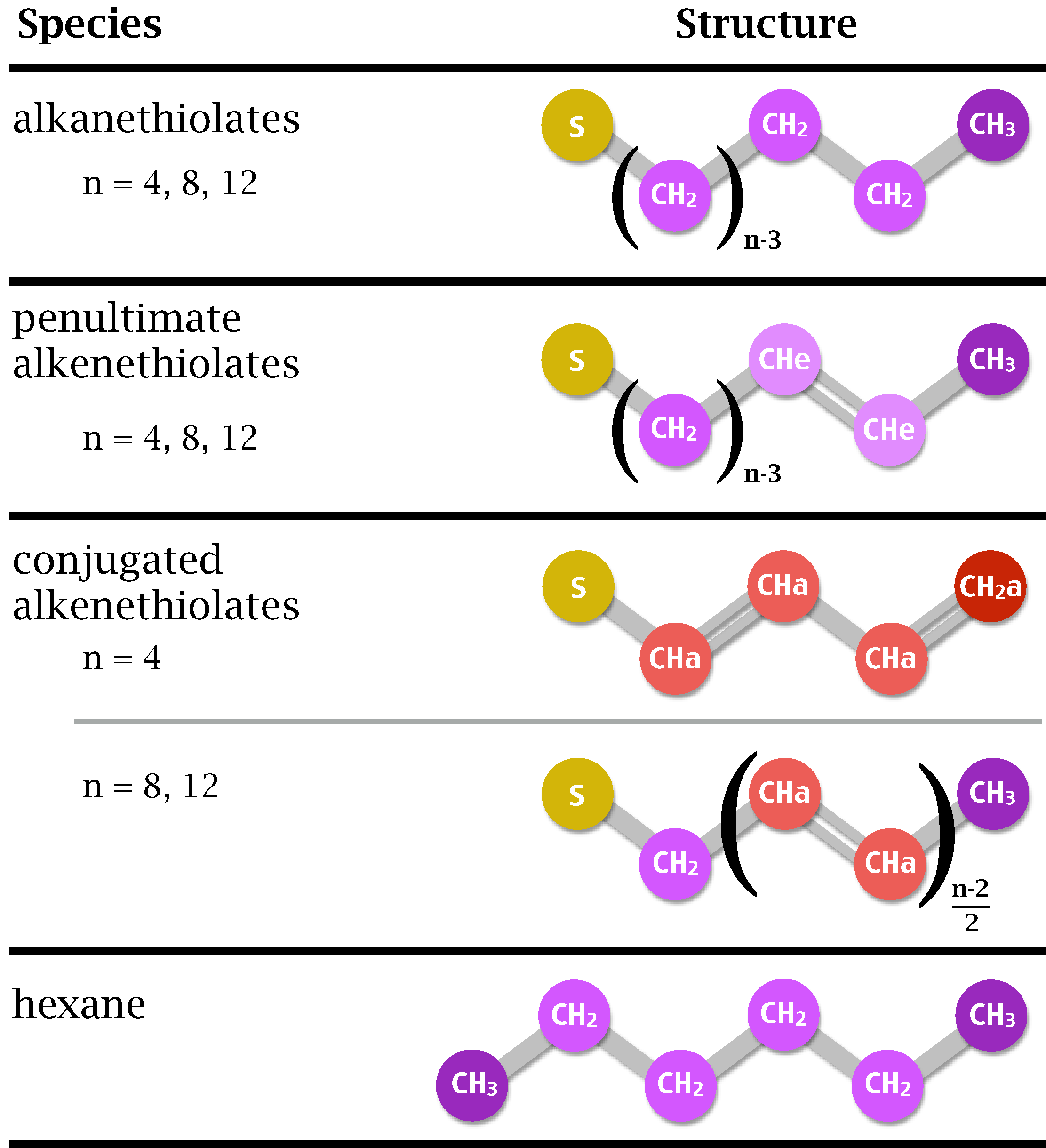}
  \caption{Topologies of the thiolate capping agents and solvent
    utilized in the simulations. The chemically-distinct sites (S,
    \ce{CH2}, \ce{CH3}, CHe, CHa and \ce{CH2a}) are treated as united
    atoms. Most parameters are taken from references
    \bibpunct{}{}{,}{n}{}{,} \protect\cite{TraPPE-UA.alkanes},
    \protect\cite{TraPPE-UA.alkylbenzenes}
    \protect\cite{TraPPE-UA.thiols}. Cross-interactions with the Au
    atoms were adapted from references
    \protect\cite{landman:1998},~\protect\cite{vlugt:cpc2007154},~and
    \protect\cite{hautman:4994}.}
  \label{fig:structures}
\bibpunct{[}{]}{,}{n}{}{,}
\end{figure}

In this work, thiolated gold nanospheres were modeled using a united
atom force field and non-equilibrium molecular dynamics. Gold
nanoparticles with radii ranging from 10 - 25 \AA\ were created from a
bulk fcc lattice.  These particles were passivated with a 50\%
coverage (compared with the coverage densities reported by Badia
\textit{et al.}) of a selection of thiolates.  Three straight-chain
thiolates of varying chain lengths and rigidities were utilized.
These are summarized in Fig. \ref{fig:structures}.  The passivated
particles were then solvated in hexane.  Details on the united atom
force field are given below and in the supporting information.\cite{supplemental}

\section{Computational Details}

\subsection{Creating a thermal flux between particles and solvent}

The non-periodic variant of the velocity shearing and scaling RNEMD
algorithm (VSS-RNEMD)\cite{Stocker:2014qq} applies a series of
velocity scaling and shearing moves at regular intervals to impose a
flux between two concentric spherical regions. To impose a thermal
flux between the shells (without an accompanying angular shear), we
solve for scaling coefficients $a$ and $b$,
\begin{eqnarray}
	a = \sqrt{1 - \frac{q_r \Delta t}{K_a - K_a^\mathrm{rot}}}\\ \nonumber\\
	b = \sqrt{1 + \frac{q_r \Delta t}{K_b - K_b^\mathrm{rot}}}
\end{eqnarray}
at each time interval.  These scaling coefficients conserve total
kinetic energy and angular momentum subject to an imposed heat rate,
$q_r$.  The coefficients also depend on the instantaneous kinetic
energy, $K_{\{a,b\}}$, and the total rotational kinetic energy of each
shell, $K_{\{a,b\}}^\mathrm{rot} = \sum_i m_i \left( \mathbf{v}_i
  \times \mathbf{r}_i \right)^2 / 2$.

The scaling coefficients are determined and the velocity changes are
applied at regular intervals, 
\begin{eqnarray}
	\mathbf{v}_i \leftarrow a \left ( \mathbf{v}_i - \left < \omega_a \right > \times \mathbf{r}_i \right ) + \left < \omega_a \right > \times \mathbf{r}_i~~\:\\
	\mathbf{v}_j \leftarrow b \left ( \mathbf{v}_j - \left < \omega_b \right > \times \mathbf{r}_j \right ) + \left < \omega_b \right > \times \mathbf{r}_j.
\end{eqnarray}
Here $\left < \omega_a \right > \times \mathbf{r}_i$ is the
contribution to the velocity of particle $i$ due to the overall
angular velocity of the $a$ shell. In the absence of an angular
momentum flux, the angular velocity $\left < \omega_a \right >$ of the
shell is nearly 0 and the resultant particle velocity is a nearly
linear scaling of the initial velocity by the coefficient $a$ or $b$.

Repeated application of this thermal energy exchange yields a radial
temperature profile for the solvated nanoparticles that depends
linearly on the applied heat rate, $q_r$. Similar to the behavior in
the slab geometries, the temperature profiles have discontinuities at
the interfaces between dissimilar materials.  The size of the
discontinuity depends on the interfacial thermal conductance, which is
the primary quantity of interest.

\subsection{Interfacial Thermal Conductance}

As described in earlier work,\cite{Stocker:2014qq} the thermal
conductance of each spherical shell may be defined as the inverse
Kapitza resistance of the shell. To describe the thermal conductance
of an interface of considerable thickness -- such as the ligand layers
shown here -- we can sum the individual thermal resistances of each
concentric spherical shell to arrive at the inverse of the total
interfacial thermal conductance. In slab geometries, the intermediate
temperatures cancel, but for concentric spherical shells, the
intermediate temperatures and surface areas remain in the final sum,
requiring the use of a series of individual resistance terms:

\begin{equation}
  \frac{1}{G} = R_\mathrm{total} = \frac{1}{q_r} \sum_i \left(T_{i+i} -
    T_i\right) 4 \pi r_i^2.
\end{equation}

The longest ligand considered here is in excess of 15 \AA\ in length,
and we use 10 concentric spherical shells to describe the total
interfacial thermal conductance of the ligand layer.

\subsection{Force Fields}

Throughout this work, gold -- gold interactions are described by the
quantum Sutton-Chen (QSC) model.\cite{Qi:1999ph} Previous
work\cite{Kuang:2011ef} has demonstrated that the electronic
contributions to heat conduction (which are missing from the QSC
model) across heterogeneous metal / non-metal interfaces are
negligible compared to phonon excitation, which is captured by the
classical model. The hexane solvent is described by the TraPPE united
atom model,\cite{TraPPE-UA.alkanes} where sites are located at the
carbon centers for alkyl groups. The TraPPE-UA model for hexane
provides both computational efficiency and reasonable accuracy for
bulk thermal conductivity values. Bonding interactions were used for
intra-molecular sites closer than 3 bonds. Effective Lennard-Jones
potentials were used for non-bonded interactions.

The TraPPE-UA force field includes parameters for thiol
molecules\cite{TraPPE-UA.thiols} as well as unsaturated and aromatic
carbon sites.\cite{TraPPE-UA.alkylbenzenes} These were used for the
thiolate molecules in our simulations, and missing parameters for the
ligands were supplemented using fits described in the supporting
information.\cite{supplemental}  Bonds are rigid in TraPPE-UA, so although equilibrium
bond distances were taken from this force field, flexible bonds were
implemented using bond stretching spring constants adapted from the
OPLS-AA force field.\cite{Jorgensen:1996sf}

To derive suitable parameters for the thiolates adsorbed on Au(111)
surfaces, we adopted the S parameters from Luedtke and
Landman\cite{landman:1998} and modified the parameters for the CTS
atom to maintain charge neutrality in the molecule.

Other interactions between metal (Au) and non-metal atoms were adapted
from an adsorption study of alkyl thiols on gold surfaces by Vlugt,
\textit{et al.}\cite{vlugt:cpc2007154} They fit an effective pair-wise
Lennard-Jones form of potential parameters for the interaction between
Au and pseudo-atoms CH$_x$ and S based on a well-established and
widely-used effective potential of Hautman and Klein for the Au(111)
surface.\cite{hautman:4994}

All additional terms to represent thiolated alkenes and conjugated
ligand moieties were parameterized as part of this work and are
available in the supporting information.\cite{supplemental}  
All simulations were carried out with the open source molecular 
dynamics package, OpenMD.\cite{openmd,OOPSE}

\subsection{Simulation Protocol}

Gold nanospheres with radii ranging from 10 - 25 \AA\ were created
from a bulk fcc lattice and were thermally equilibrated prior to the
addition of ligands. A 50\% coverage of ligands (based on coverages
reported by Badia, \textit{et al.}\cite{Badia1996:2}) was placed on
the surface of the equilibrated nanoparticles using
Packmol\cite{packmol}. We have chosen three lengths for the
straight-chain ligands, $C_4$, $C_8$, and $C_{12}$, differentiated by
the number of carbons in the chains.  Additionally, to explore the
effects of ligand flexibility, we have used three levels of ligand
``stiffness''.  The most flexible chain is a fully saturated
alkanethiolate, while moderate rigidity is introduced using an alkene
thiolate with one double bond in the penultimate (solvent-facing)
carbon-carbon location.  The most rigid ligands are fully-conjugated
chains where all of the carbons are represented with conjugated (aryl)
united-atom carbon atoms (CHar or terminal \ce{CH2ar}).

\begin{figure}
  \includegraphics[width=\linewidth]{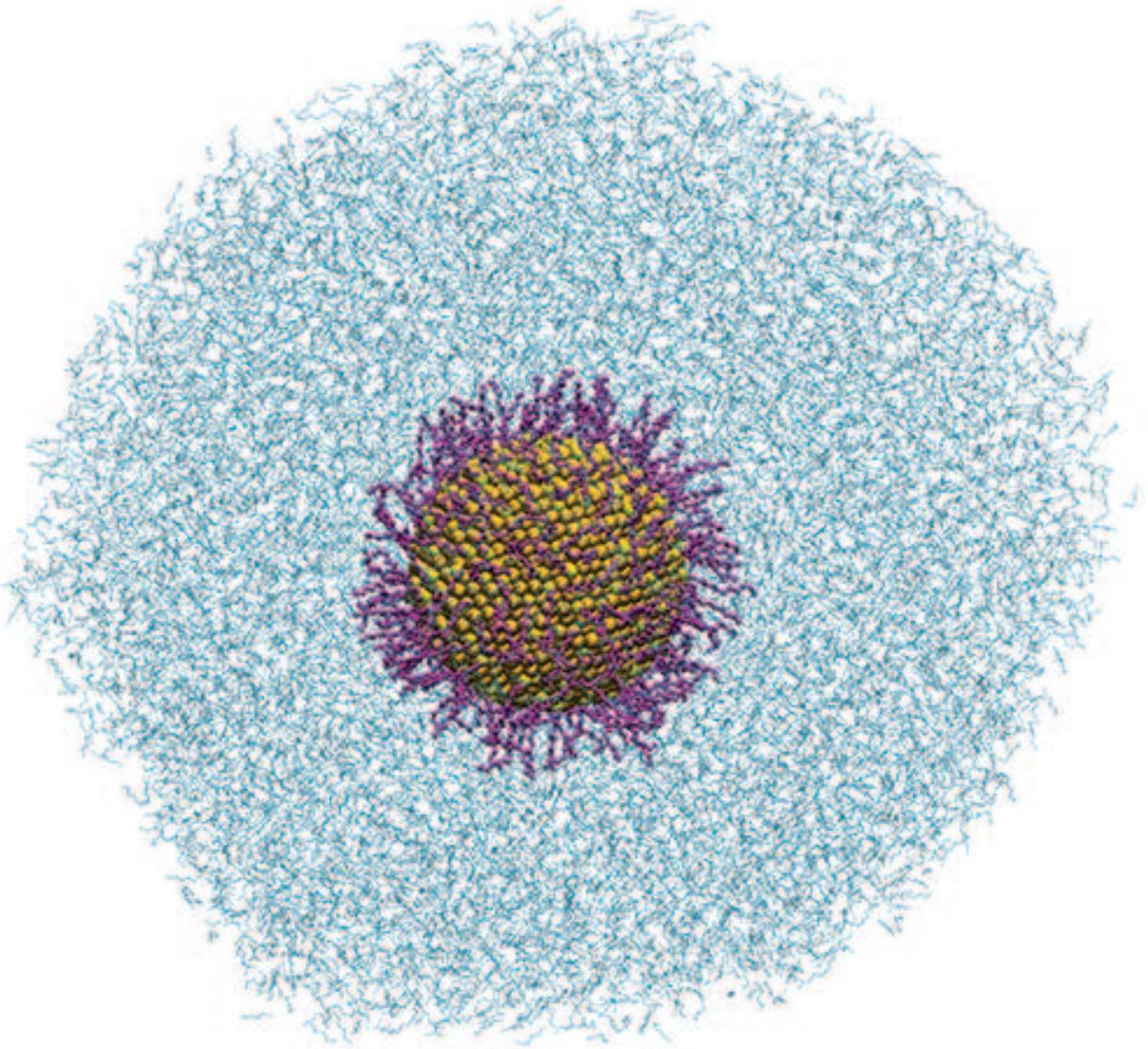}
  \caption{A 25 \AA\ radius gold nanoparticle protected with a
    half-monolayer of TraPPE-UA dodecanethiolate (C$_{12}$) ligands
    and solvated in TraPPE-UA hexane. The interfacial thermal
    conductance is computed by applying a kinetic energy flux between
    the nanoparticle and an outer shell of solvent.}
  \label{fig:NP25_C12h1}
\end{figure}

The nanoparticle / ligand complexes were thermally equilibrated to
allow for ligand conformational flexibility. Packmol was then used to
solvate the structures inside a spherical droplet of hexane. The
thickness of the solvent layer was chosen to be at least 1.5$\times$
the combined radius of the nanoparticle / ligand structure. The fully
solvated system was equilibrated for at least 1 ns using the
``Langevin Hull'' algorithm to apply 50 atm of pressure and a target
temperature of 250 K.\cite{Vardeman2011} Typical system sizes ranged
from 18,310 united atom sites for the 10 \AA\ particles with $C_4$
ligands to 89,490 sites for the 25 \AA\ particles with $C_{12}$
ligands.  Figure \ref{fig:NP25_C12h1} shows one of the solvated 25
\AA\ nanoparticles passivated with the $C_{12}$ alkane thiolate
ligands.

\begin{figure}
  \includegraphics[width=\linewidth]{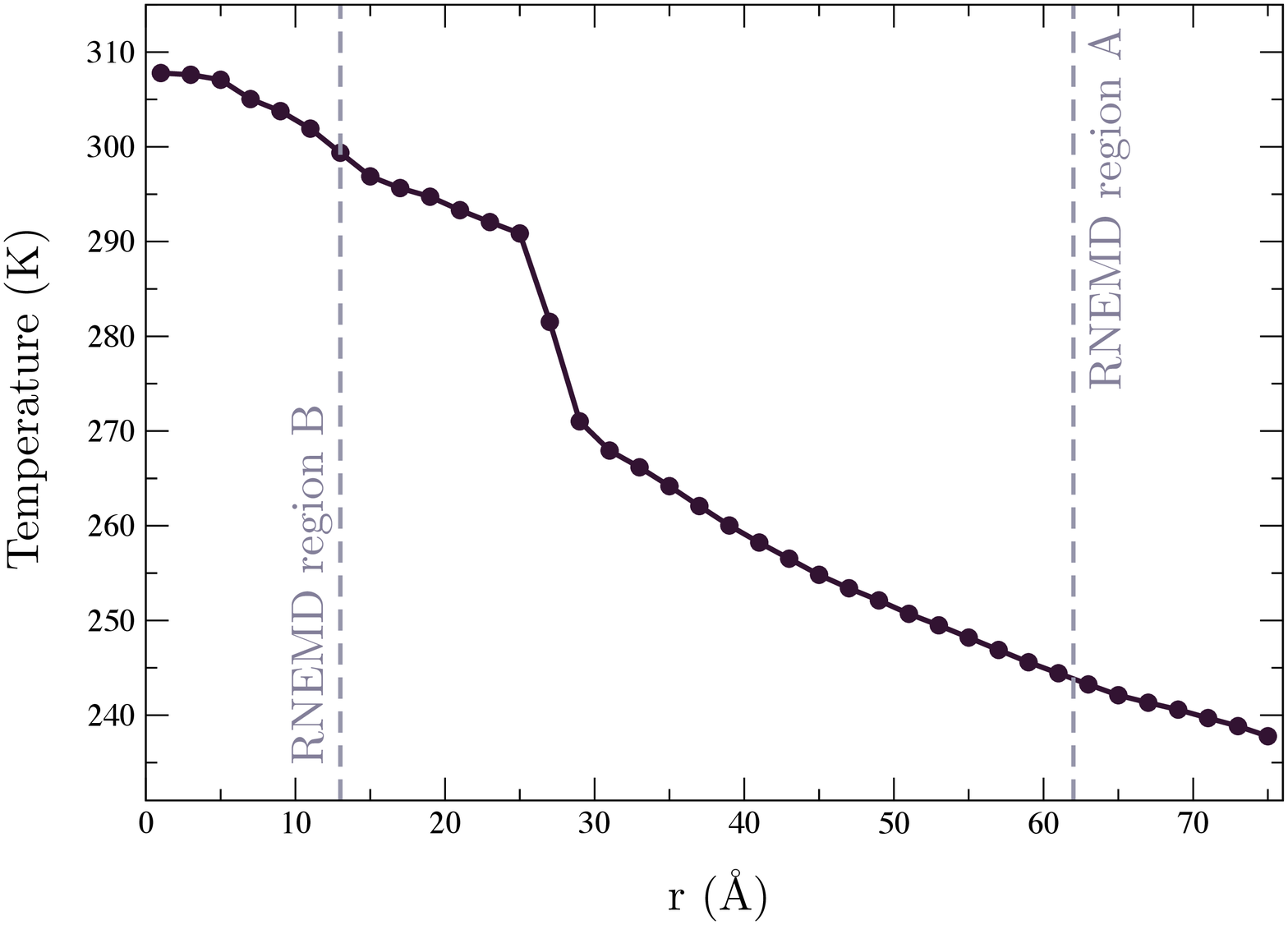}
  \caption{Radial temperature profile for a 25 \AA\ radius particle
    protected with a 50\% coverage of TraPPE-UA butanethiolate (C$_4$)
    ligands and solvated in TraPPE-UA hexane. A kinetic energy flux is
    applied between RNEMD region A and RNEMD region B. The size of the
    temperature discontinuity at the interface is governed by the
    interfacial thermal conductance.}
	\label{fig:temp_profile}
\end{figure}

Once equilibrated, thermal fluxes were applied for 1 ns, until stable
temperature gradients had developed (see figure
\ref{fig:temp_profile}). Systems were run under moderate pressure (50
atm) with an average temperature (250K) that maintained a compact
solvent cluster and avoided formation of a vapor layer near the heated
metal surface.  Pressure was applied to the system via the
non-periodic ``Langevin Hull'' algorithm.\cite{Vardeman2011} However,
thermal coupling to the external temperature bath was removed to avoid
interference with the imposed RNEMD flux.

Although the VSS-RNEMD moves conserve \emph{total} angular momentum
and energy, systems which contain a metal nanoparticle embedded in a
significant volume of solvent will still experience nanoparticle
diffusion inside the solvent droplet. To aid in measuring an accurate
temperature profile for these systems, a single gold atom at the
origin of the coordinate system was assigned a mass $10,000 \times$
its original mass. The bonded and nonbonded interactions for this atom
remain unchanged and the heavy atom is excluded from the RNEMD
velocity scaling.  The only effect of this gold atom is to effectively
pin the nanoparticle at the origin of the coordinate system, thereby
preventing translational diffusion of the nanoparticle due to Brownian
motion.

To provide statistical independence, five separate configurations were
simulated for each particle radius and ligand. The structures were
unique, starting at the point of ligand placement, in order to sample
multiple surface-ligand configurations.

\section{Results}

\begin{figure}
  \includegraphics[width=\linewidth]{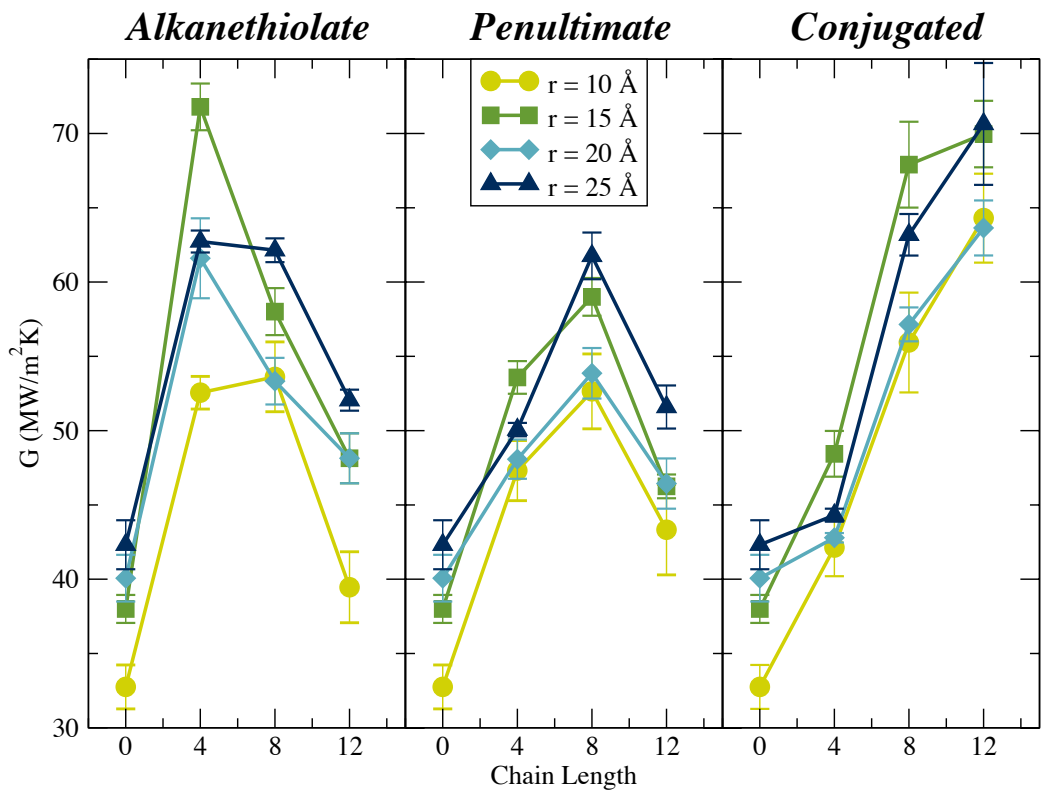}
  \caption{Interfacial thermal conductance ($G$) values for 4 sizes of
    solvated nanoparticles that are bare or protected with a 50\%
    coverage of C$_{4}$, C$_{8}$, or C$_{12}$ thiolate
    ligands. Ligands of different flexibility are shown in separate
    panels.  The middle panel indicates ligands which have a single
    carbon-carbon double bond in the penultimate position.}
  \label{fig:NPthiols_G}
\end{figure}

We modeled four sizes of nanoparticles ($R =$ 10, 15, 20, and 25
\AA). The smallest particle size produces the lowest interfacial
thermal conductance values for most of the of protecting groups
(Fig. \ref{fig:NPthiols_G}).  Between the other three sizes of
nanoparticles, there is no systematic dependence of the interfacial
thermal conductance on the nanoparticle size. It is likely that the
differences in local curvature of the nanoparticle sizes studied here
do not disrupt the ligand packing and behavior in drastically
different ways.

Unlike our previous study of varying thiolate ligand chain lengths on
planar Au(111) surfaces, the interfacial thermal conductance of
ligand-protected nanospheres exhibits a distinct dependence on the
ligand identity. A half-monolayer coverage of ligands yields
interfacial conductance that is strongly dependent on both ligand
length and flexibility.

There are many factors that could be playing a role in the
ligand-dependent conductuance.  The sulfur-gold interaction is
particularly strong, and the presence of the ligands can easily
disrupt the crystalline structure of the gold at the surface of the
particles, providing more efficient scattering of phonons into the
ligand / solvent layer. This effect would be particularly important at
small particle sizes.

In previous studies of mixed-length ligand layers with full coverage,
we observed that ligand-solvent alignment was an important factor for
heat transfer into the solvent.  With high surface curvature and lower
effective coverages, ligand behavior also becomes more complex. Some
chains may be lying down on the surface, and solvent may not be
penetrating the ligand layer to the same degree as in the planar
surfaces.  

Additionally, the ligand flexibility directly alters the vibrational
density of states for the layer that mediates the transfer of phonons
between the metal and the solvent. This could be a partial explanation
for the observed differences between the fully conjugated and more
flexible ligands.

In the following sections we provide details on how we
measure surface corrugation, solvent-ligand interpenetration, and
ordering of the solvent and ligand at the surfaces of the
nanospheres.  We also investigate the overlap between vibrational
densities of states for the various ligands.

\subsection{Corrugation of the Particle Surface}

The bonding sites for thiols on gold surfaces have been studied
extensively and include configurations beyond the traditional atop,
bridge, and hollow sites found on planar surfaces. In particular, the
deep potential well between the gold atoms and the thiolate sulfur
atoms leads to insertion of the sulfur into the gold lattice and
displacement of interfacial gold atoms. The degree of ligand-induced
surface restructuring may have an impact on the interfacial thermal
conductance and is an important phenomenon to quantify.

Henz, \textit{et al.}\cite{Henz:2008qf} used the metal
density as a function of radius to measure the degree of mixing
between the thiol sulfurs and surface gold atoms at the edge of a
nanoparticle. Although metal density is important, disruption of the
local crystalline ordering would also have a large effect on the
phonon spectrum in the particles. To measure this effect, we use the
fraction of gold atoms exhibiting local fcc ordering as a function of
radius to describe the ligand-induced disruption of the nanoparticle
surface.

The local bond orientational order can be described using the method
of Steinhardt \textit{et al.}\cite{Steinhardt1983} The local bonding
environment, $\bar{q}_{\ell m}$, for each atom in the system is
determined by averaging over the spherical harmonics between that atom
and each of its neighbors,
\begin{equation}
\bar{q}_{\ell m} = \sum_i Y_\ell^m(\theta_i, \phi_i)
\end{equation}
where $\theta_i$ and $\phi_i$ are the relative angular coordinates of
neighbor $i$ in the laboratory frame.  A global average orientational
bond order parameter, $\bar{Q}_{\ell m}$, is the average over each
$\bar{q}_{\ell m}$ for all atoms in the system. To remove the
dependence on the laboratory coordinate frame, the third order
rotationally invariant combination of $\bar{Q}_{\ell m}$,
$\hat{w}_\ell$, is utilized here.\cite{Steinhardt1983,Vardeman:2008fk}

For $\ell=4$, the ideal face-centered cubic (fcc), body-centered cubic
(bcc), hexagonally close-packed (hcp), and simple cubic (sc) local
structures exhibit $\hat{w}_4$ values of -0.159, 0.134, 0.159, and
0.159, respectively. Because $\hat{w}_4$ exhibits an extreme value for
fcc structures, it is ideal for measuring local fcc
ordering. The spatial distribution of $\hat{w}_4$ local bond
orientational order parameters, $p(\hat{w}_4 , r)$, can provide
information about the location of individual atoms that are central to
local fcc ordering.

The fraction of fcc-ordered gold atoms at a given radius in the
nanoparticle,
\begin{equation}
	f_\mathrm{fcc}(r) = \int_{-\infty}^{w_c} p(\hat{w}_4, r) d \hat{w}_4
\end{equation}
is described by the distribution of the local bond orientational order
parameters, $p(\hat{w}_4, r)$, and $w_c$, a cutoff for the peak
$\hat{w}_4$ value displayed by fcc structures. A $w_c$ value of -0.12
was chosen to isolate the fcc peak in $\hat{w}_4$.

\begin{figure}
  \includegraphics[width=\linewidth]{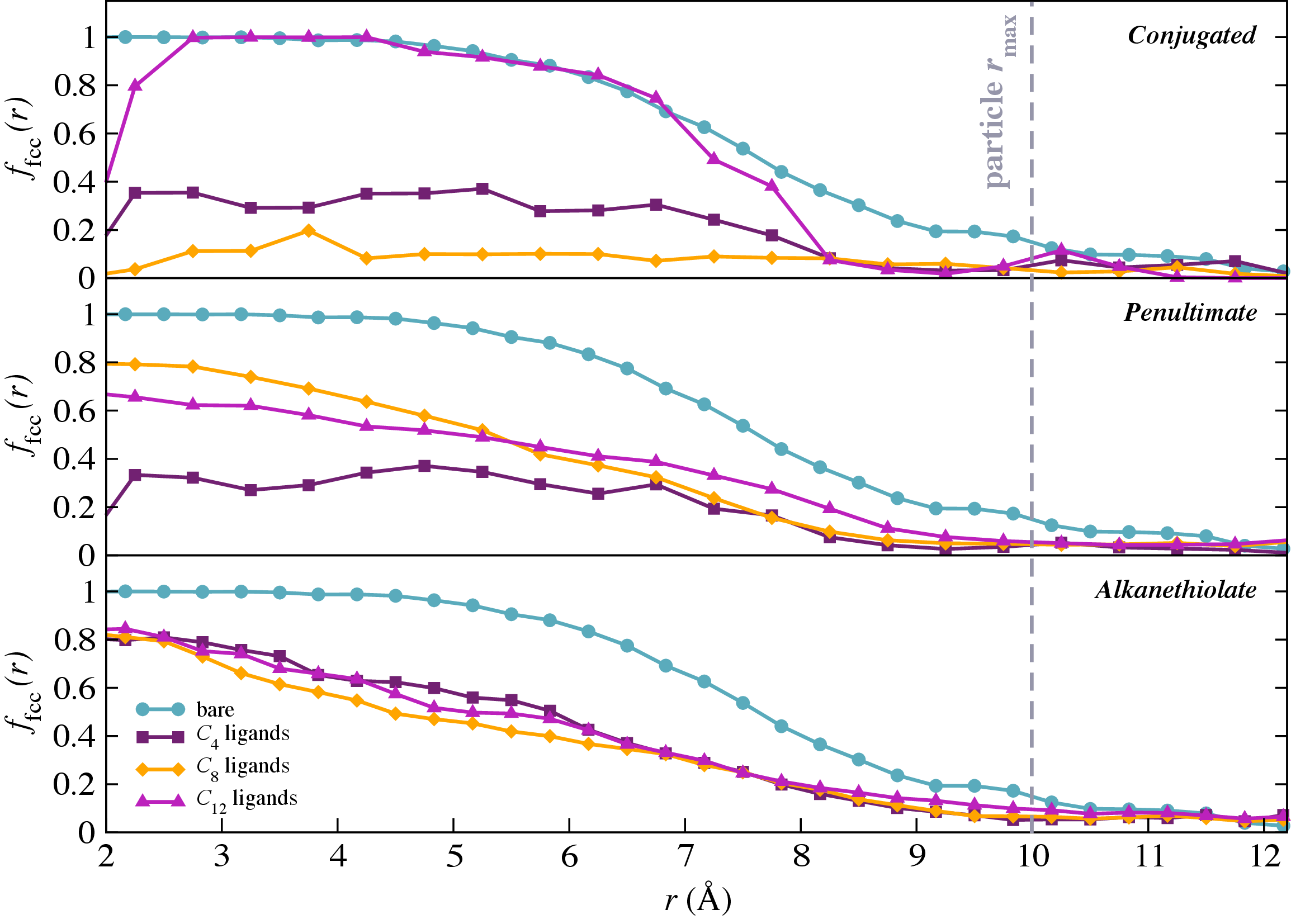}
  \caption{Fraction of gold atoms with fcc ordering as a function of
    radius for a 10 \AA\ radius nanoparticle. The decreased fraction
    of fcc-ordered atoms in ligand-protected nanoparticles relative to
    bare particles indicates restructuring of the nanoparticle surface
    by the thiolate sulfur atoms.}
  \label{fig:Corrugation}
\end{figure}

As illustrated in Figure \ref{fig:Corrugation}, the presence of
ligands decreases the fcc ordering of the gold atoms at the
nanoparticle surface. For the smaller nanoparticles, this disruption
extends into the core of the nanoparticle, indicating widespread
disruption of the lattice.

We may describe the thickness of the disrupted nanoparticle surface by
defining a corrugation factor, $c$, as the ratio of the radius at
which the fraction of gold atoms with fcc ordering is 0.9 and the
radius at which the fraction is 0.5.

\begin{equation}
	c = 1 - \frac{r(f_\mathrm{fcc} = 0.9)}{r(f_\mathrm{fcc} = 0.5)}
\end{equation}

A sharp interface will have a steep drop in $f_\mathrm{fcc}$ at the
edge of the particle ($c \rightarrow$ 0). In the opposite limit where
the entire nanoparticle surface is restructured by ligands, the radius
at which there is a high probability of fcc ordering moves
dramatically inward ($c \rightarrow$ 1).

\begin{figure}
  \includegraphics[width=\linewidth]{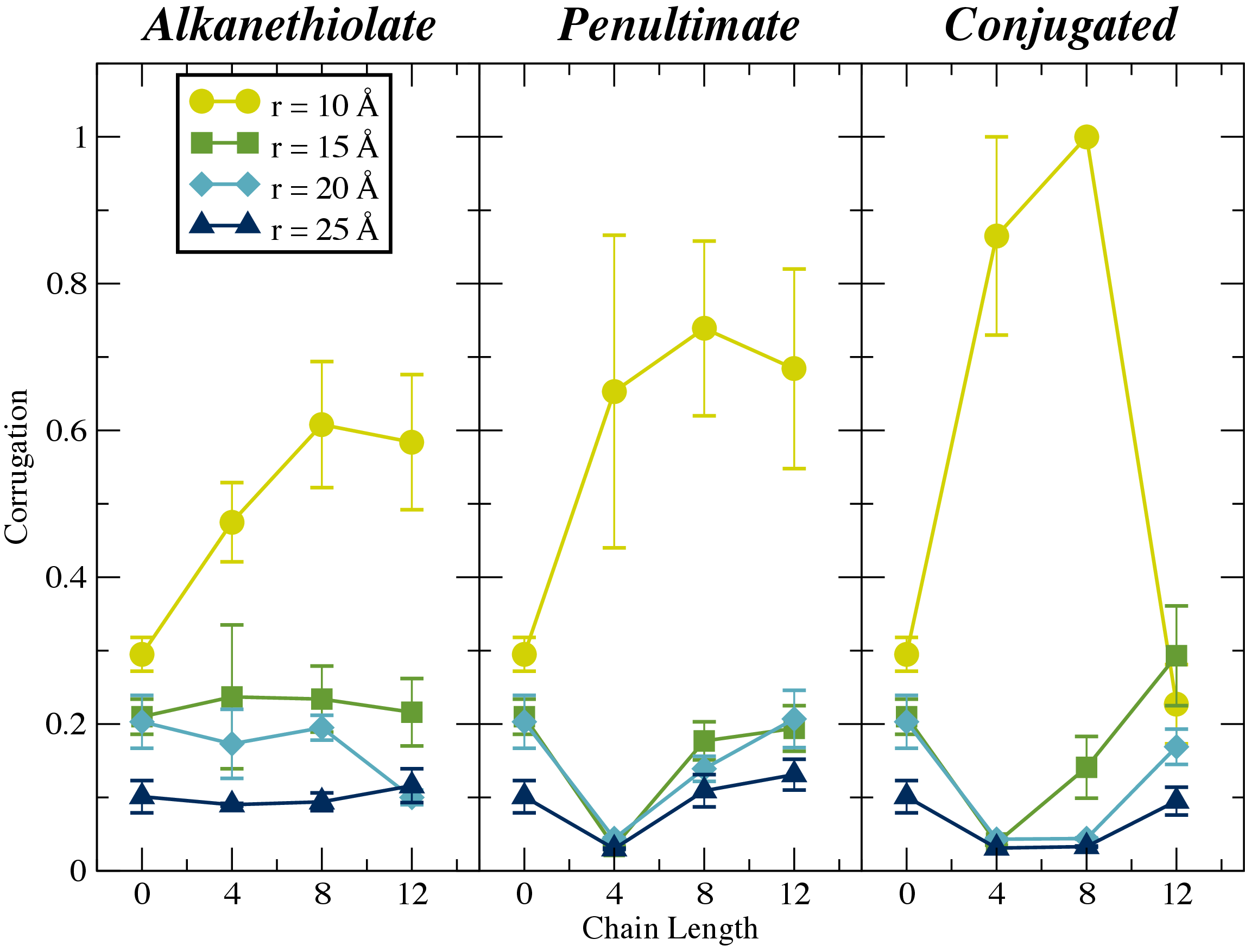}
  \caption{Computed corrugation values for 4 sizes of solvated
    nanoparticles that are bare or protected with a 50\% coverage of
    C$_{4}$, C$_{8}$, or C$_{12}$ thiolate ligands.  The smallest (10
    \AA ) particles show significant disruption to their crystal
    structures, and the length and stiffness of the ligands is a
    contributing factor to the surface disruption.}
  \label{fig:NPthiols_corrugation}
\end{figure}

The computed corrugation factors are shown in Figure
\ref{fig:NPthiols_corrugation} for bare nanoparticles and for
ligand-protected particles as a function of ligand chain length. The
largest nanoparticles are only slightly restructured by the presence
of ligands on the surface, while the smallest particle ($r$ = 10 \AA)
exhibits significant disruption of the original fcc ordering when
covered with a half-monolayer of thiol ligands.

Because the thiolate ligands do not significantly alter the larger
particle crystallinity, the surface corrugation does not seem to be a
likely candidate to explain the large increase in thermal conductance
at the interface when ligands are added.

\subsection{Orientation of Ligand Chains}

\begin{figure}
  \includegraphics[width=\linewidth]{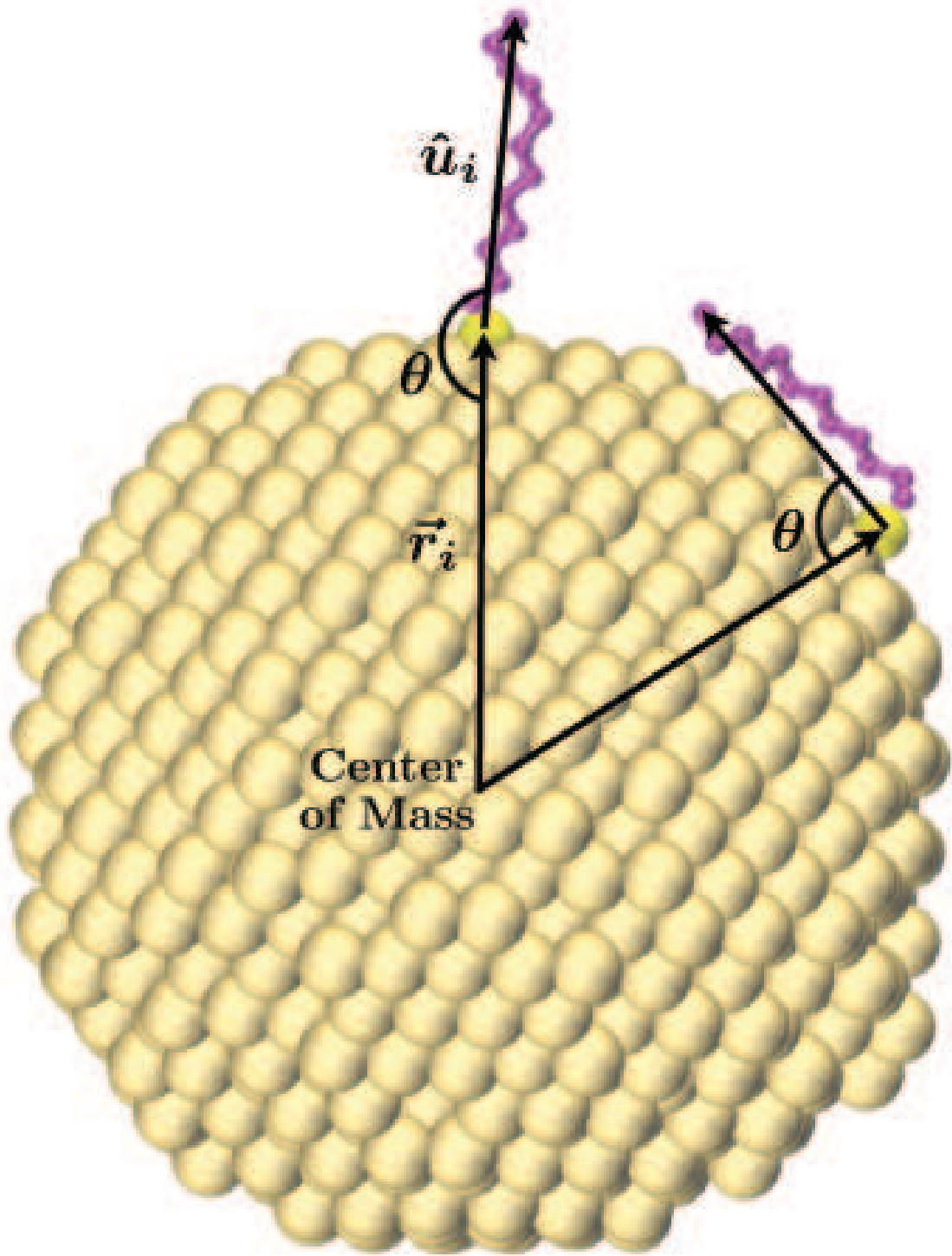}
  \caption{The two extreme cases of ligand orientation relative to the
    nanoparticle surface: the ligand completely outstretched
    ($\cos{(\theta)} = -1$) and the ligand fully lying down on the
    particle surface ($\cos{(\theta)} = 0$).}
  \label{fig:NP_pAngle}
\end{figure}

Previous theoretical work on heat conduction through alkane chains has
shown that short chains are dominated by harmonic interactions, where
the energy is carried ballistically through the
chain.\cite{Segal:2003qy} As the saturated ligand chain length
increases in length, it exhibits significantly more conformational
flexibility. Thus, different lengths of ligands should favor different
chain orientations on the surface of the nanoparticle, and can
localize the ligand vibrational density of states close to the
particle, lowering the effectiveness of the heat
conduction.\cite{Segal:2003qy} To determine the distribution of ligand
orientations relative to the particle surface we examine the
probability of finding a ligand with a particular orientation relative
to the surface normal of the nanoparticle,
\begin{equation} 
\cos{(\theta)}=\frac{\vec{r}_i\cdot\hat{u}_i}{|\vec{r}_i||\hat{u}_i|}
\end{equation}
where $\vec{r}_{i}$ is the vector between the cluster center of mass
and the sulfur atom on ligand molecule {\it i}, and $\hat{u}_{i}$ is
the  vector between the sulfur atom and \ce{CH3} pseudo-atom on ligand
molecule {\it i}. As depicted in Figure \ref{fig:NP_pAngle}, $\theta
\rightarrow 180^{\circ}$ for a ligand chain standing upright on the
particle ($\cos{(\theta)} \rightarrow -1$) and $\theta \rightarrow
90^{\circ}$ for a ligand chain lying down on the surface
($\cos{(\theta)} \rightarrow 0$). As the thiolate alkane chain
increases in length and becomes more flexible, the ligands are more
willing to lie down on the nanoparticle surface and exhibit increased
population at $\cos{(\theta)} = 0$.

An order parameter describing the average ligand chain orientation relative to
the nanoparticle surface is available using the second order Legendre
parameter,
\begin{equation}
	P_2 = \left< \frac{1}{2} \left(3\cos^2(\theta) - 1 \right) \right>
\end{equation}

\begin{figure}
  \includegraphics[width=\linewidth]{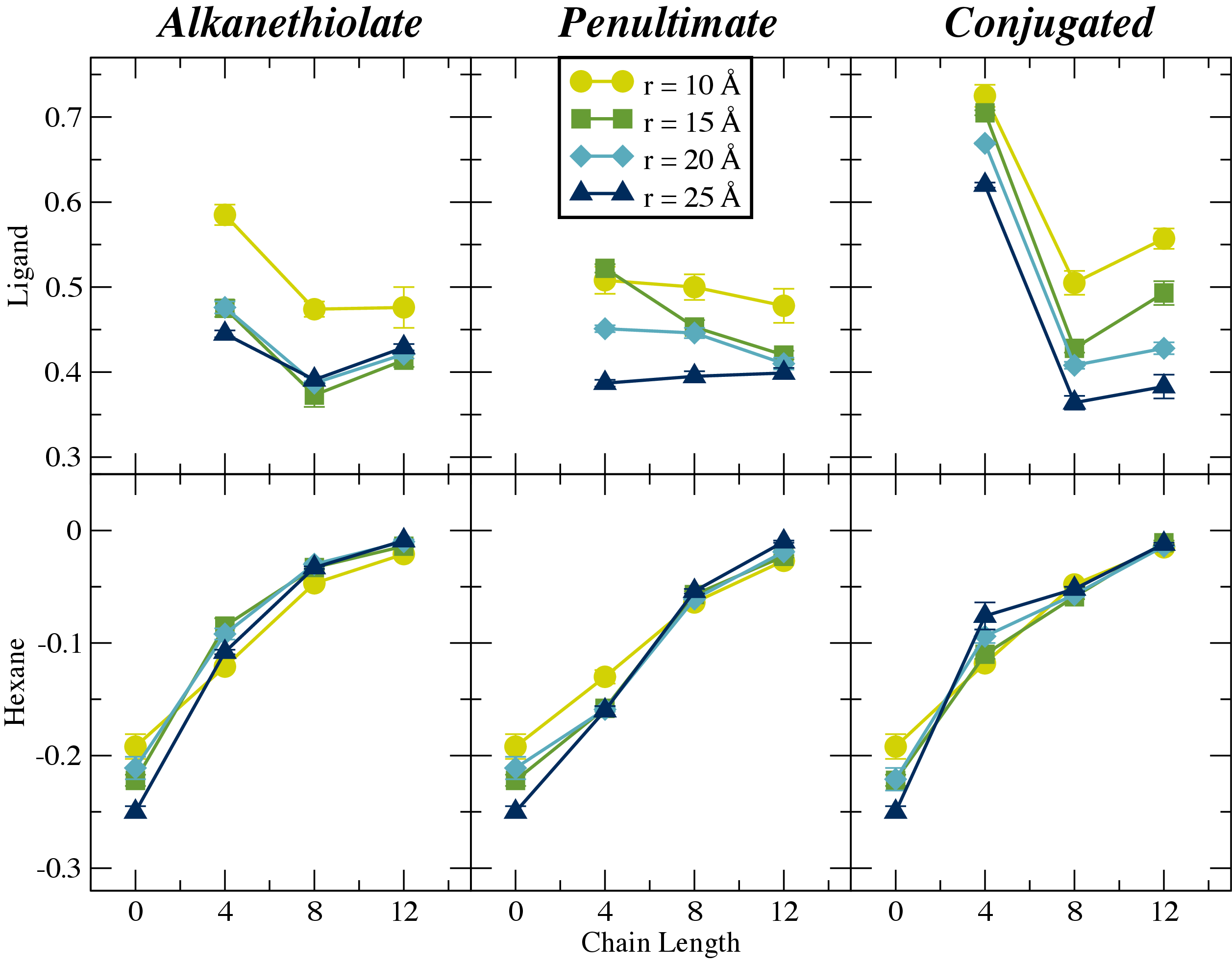}
  \caption{Computed ligand and interfacial solvent orientational $P_2$
    values for 4 sizes of solvated nanoparticles that are bare or
    protected with a 50\% coverage of C$_{4}$, C$_{8}$, or C$_{12}$
    alkanethiolate ligands. Increasing stiffness of the ligand orients
    these molecules normal to the particle surface, while the length
    of the ligand chains works to prevent solvent from lying flat on
    the surface.}
  \label{fig:NPthiols_P2}
\end{figure}

Ligand populations that are perpendicular to the particle surface have
$P_2$ values of 1, while ligand populations lying flat on the
nanoparticle surface have $P_2$ values of $-0.5$. Disordered ligand
layers will exhibit mean $P_2$ values of 0. As shown in Figure
\ref{fig:NPthiols_P2} the ligand $P_2$ values approaches 0 as
ligand chain length -- and ligand flexibility -- increases.

\subsection{Orientation of Interfacial Solvent}

Similarly, we examined the distribution of \emph{hexane} molecule
orientations relative to the particle surface using the same angular
analysis utilized for the ligand chain orientations. In this case,
$\vec{r}_i$ is the vector between the particle center of mass and one
of the \ce{CH2} pseudo-atoms in the middle of hexane molecule $i$ and
$\hat{u}_i$ is the vector between the two \ce{CH3} pseudo-atoms on
molecule $i$. Since we are only interested in the orientation of
solvent molecules near the ligand layer, we select only the hexane
molecules within a specific $r$-range, between the edge of the
particle and the end of the ligand chains. A large population of
hexane molecules with $\cos{(\theta)} \sim \pm 1$ would indicate
interdigitation of the solvent molecules between the upright ligand
chains. A more random distribution of $\cos{(\theta)}$ values
indicates a disordered arrangement of solvent molecules near the particle
surface. Again, $P_2$ order parameter values provide a population
analysis for the solvent that is close to the particle surface.

The average orientation of the interfacial solvent molecules is
notably flat on the \emph{bare} nanoparticle surfaces. This blanket of
hexane molecules on the particle surface may act as an insulating
layer, increasing the interfacial thermal resistance. As the length
(and flexibility) of the ligand increases, the average interfacial
solvent P$_2$ value approaches 0, indicating a more random orientation
of the ligand chains. The average orientation of solvent within the
$C_8$ and $C_{12}$ ligand layers is essentially random. Solvent
molecules in the interfacial region of $C_4$ ligand-protected
nanoparticles do not lie as flat on the surface as in the case of the
bare particles, but are not as randomly oriented as the longer ligand
lengths.

These results are particularly interesting in light of our previous
results\cite{Stocker:2013cl}, where solvent molecules readily filled
the vertical gaps between neighboring ligand chains and there was a
strong correlation between ligand and solvent molecular
orientations. It appears that the introduction of surface curvature
and a lower ligand packing density creates a disordered ligand layer
that lacks well-formed channels for the solvent molecules to occupy.

\subsection{Solvent Penetration of Ligand Layer}

\begin{figure}
  \includegraphics[width=\linewidth]{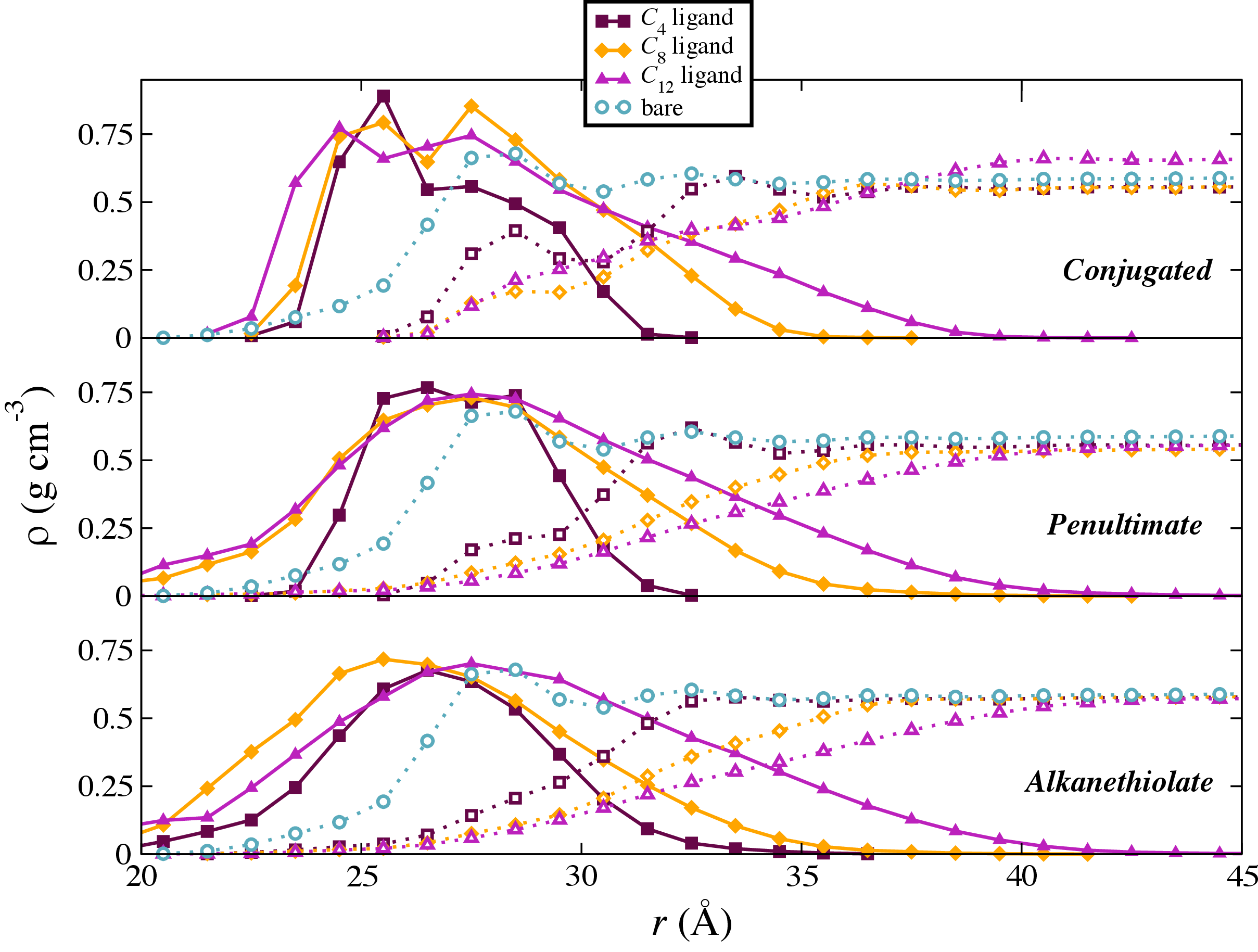}
  \caption{Radial density profiles for 25 \AA\ radius nanoparticles
    with no ligands (circles), C$_{4}$ ligands (squares), C$_{8}$
    ligands (diamonds), and C$_{12}$ ligands (triangles). Ligand
    density is indicated with filled symbols, solvent (hexane) density
    is indicated with open symbols. As ligand chain length increases,
    the nearby solvent is excluded from the ligand layer.  The
    conjugated ligands (upper panel) can create a separated solvent
    shell within the ligand layer and also allow significantly more
    solvent to penetrate close to the particle.}
  \label{fig:density}
\end{figure}

The extent of ligand -- solvent interaction is also determined by the
degree to which these components occupy the same region of space
adjacent to the nanoparticle. The radial density profiles of these
components help determine this degree of interaction.  Figure
\ref{fig:density} shows representative density profiles for solvated
25 \AA\ radius nanoparticles with no ligands, and with a 50\% coverage
of C$_{4}$, C$_{8}$, and C$_{12}$ thiolates.

The differences between the radii at which the hexane surrounding the
ligand-covered particles reaches bulk density correspond nearly
exactly to the differences between the lengths of the ligand
chains. Beyond the edge of the ligand layer, the solvent reaches its
bulk density within a few angstroms. The differing shapes of the
density curves indicate that the solvent is increasingly excluded from
the ligand layer as the chain length increases.

The conjugated ligands create a distinct solvent shell within the
ligand layer and also allow significantly more solvent to penetrate
close to the particle.  We define a density overlap parameter,
\begin{equation}
O_{l-s} = \frac{1}{V} \int_0^{r_\mathrm{max}} 4 \pi r^2 \frac{4 \rho_l(r) \rho_s(r)}{\left(\rho_l(r) +
    \rho_s(r)\right)^2} dr
\end{equation}
where $\rho_l(r)$ and $\rho_s(r)$ are the ligand and solvent densities
at a radius $r$, and $V$ is the total integration volume
($V = 4\pi r_\mathrm{max}^3 / 3$).  The fraction in the integrand is a
dimensionless quantity that is unity when ligand and solvent densities
are identical at radius $r$, but falls to zero when either of the two
components are excluded from that region.

\begin{figure}
  \includegraphics[width=\linewidth]{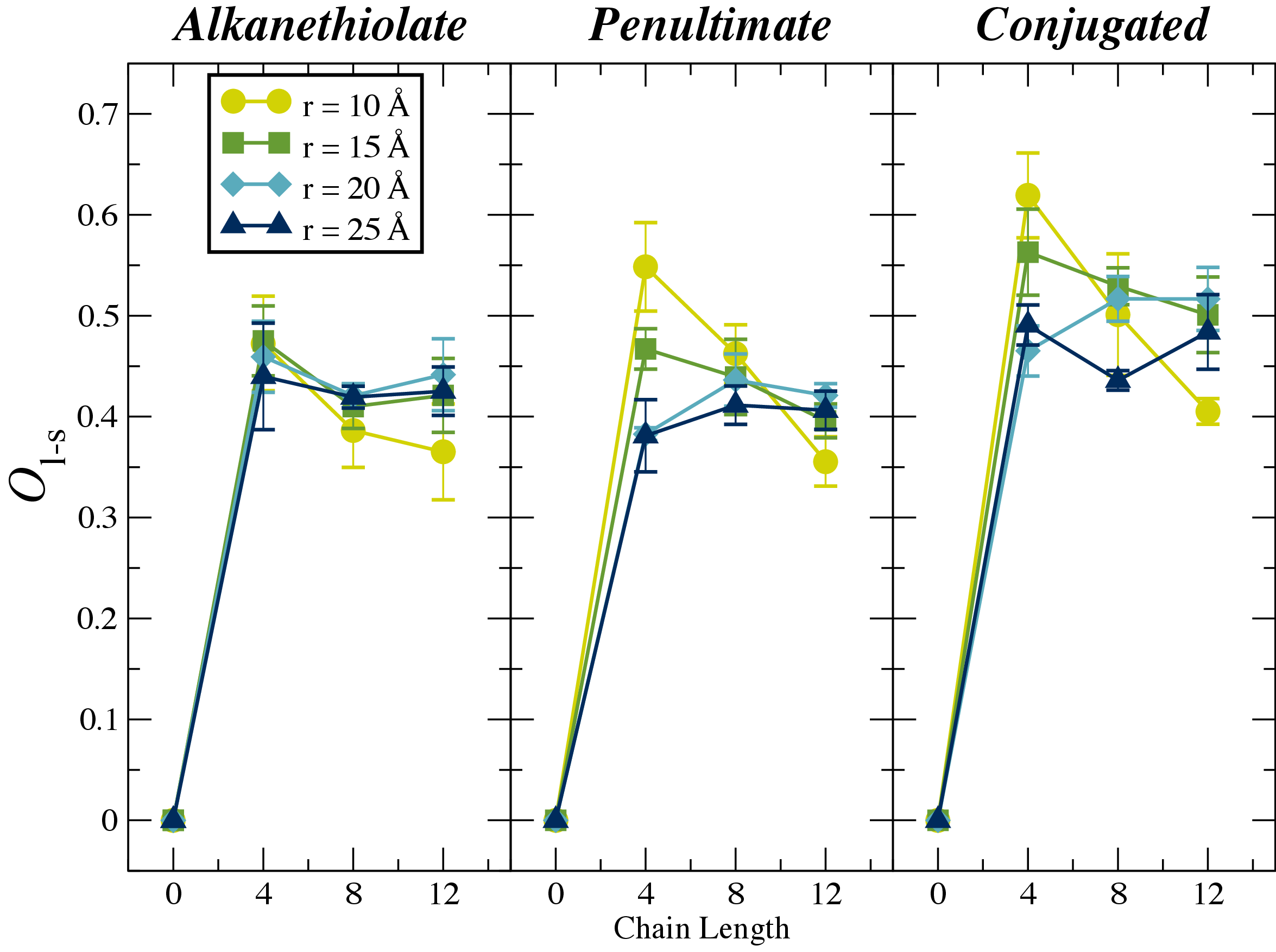}
  \caption{Density overlap parameters ($O_{l-s}$) for solvated
    nanoparticles protected by thiolate ligands. In general, the
    rigidity of the fully-conjugated ligands provides the easiest
    route for solvent to enter the interfacial region. Additionally,
    shorter chains allow a greater degree of solvent penetration of
    the ligand layer.}
  \label{fig:rho3}
\end{figure}

The density overlap parameters are shown in Fig. \ref{fig:rho3}.  The
calculated overlap parameters indicate that the conjugated ligand
allows for the most solvent penetration close to the particle, and
that shorter chains generally permit greater solvent penetration in
the interfacial region. Increasing overlap can certainly allow for
enhanced thermal transport, but this is clearly not the only
contributing factor. Even when the solvent and ligand are in close
physical contact, there must also be good vibrational overlap between
the phonon densities of states in the ligand and solvent to transmit
vibrational energy between the two materials.

\subsection{Ligand-mediated Vibrational Overlap}

In phonon scattering models for interfacial thermal
conductance,\cite{Swartz:1989uq,Young:1989xy,Cahill:2003fk,Reddy:2005fk,Schmidt:2010nr}
the frequency-dependent transmission probability
($t_{a \rightarrow b}(\omega)$) predicts phonon transfer between
materials $a$ and $b$.  Many of the models for interfacial phonon
transmission estimate this quantity using the phonon density of states
and group velocity, and make use of a Debye model for the density of
states in the solid.

A consensus picture is that in order to transfer the energy carried by
an incoming phonon of frequency $\omega$ on the $a$ side, the phonon
density of states on the $b$ side must have a phonon of the same
frequency. The overlap of the phonon densities of states, particularly
at low frequencies, therefore contributes to the transfer of heat.
Phonon scattering must also be done in a direction perpendicular to
the interface.  In the geometries described here, there are two
interfaces (particle $\rightarrow$ ligand, and ligand $\rightarrow$
solvent), and the vibrational overlap between the ligand and the other
two components is going to be relevant to heat transfer.
 
\begin{figure}
  \includegraphics[width=\linewidth]{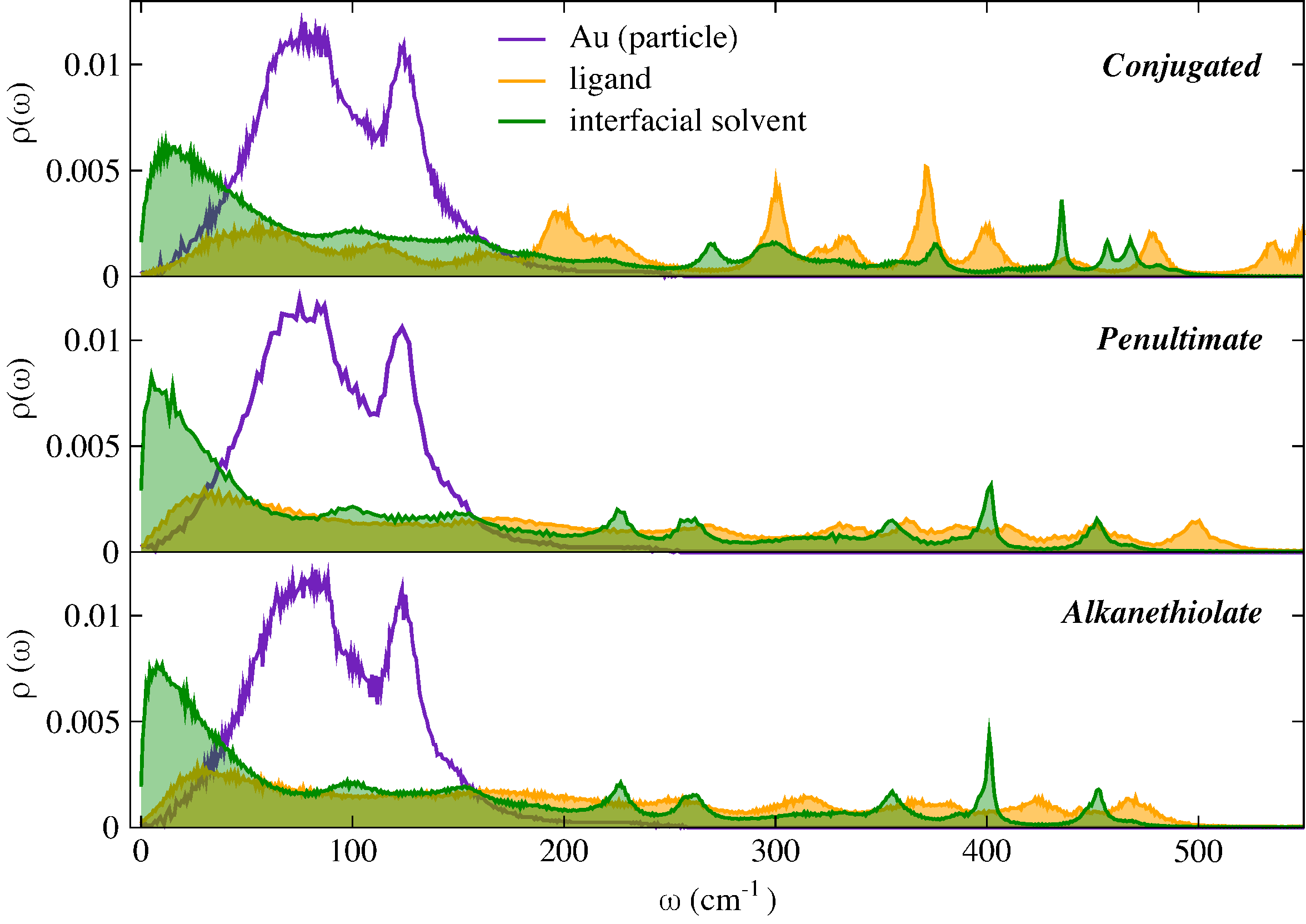}
  \caption{The low frequency portion of the vibrational density of
    states for three chemical components (gold nanoparticles, C$_{12}$
    ligands, and hexane solvent). These densities of states were
    computed using the velocity autocorrelation functions for atoms in
    the interfacial region, constructed with velocities projected onto
    a direction normal to the interface.}
  \label{fig:vdos}
\end{figure}

To estimate the relevant densities of states, we have projected the
velocity of each atom $i$ in the region of the interface onto a
direction normal to the interface. For the nanosphere geometries
studied here, the normal direction depends on the instantaneous
positon of the atom relative to the center of mass of the particle.
\begin{equation}
v_\perp(t) = \mathbf{v}(t) \cdot \frac{\mathbf{r}(t)}{\left|\mathbf{r}(t)\right|}
\end{equation}
The quantity $v_\perp(t)$ measures the instantaneous velocity of an
atom in a direction perpendicular to the nanoparticle interface.  In
the interfacial region, the autocorrelation function of these
velocities,
\begin{equation}
  C_\perp(t) = \left< v_\perp(t) \cdot v_\perp(0) \right>,
\end{equation}
will include contributions from all of the phonon modes present at the
interface.  The Fourier transform of the time-symmetrized
autocorrelation function provides an estimate of the vibrational
density of states,\cite{Shin:2010sf}
\begin{equation}
  \rho(\omega) = \frac{1}{\tau} \int_{-\tau/2}^{\tau/2} C_\perp(t) e^{-i
    \omega t} dt.
\end{equation}
Here $\tau$ is the total observation time for the autocorrelation
function.  In Fig.~\ref{fig:vdos} we show the low-frequency region of
the normalized vibrational densities of states for the three chemical
components (gold nanoparticle, C$_{12}$ ligands, and interfacial
solvent).  The double bond in the penultimate location is a small
perturbation on ligands of this size, and that is reflected in
relatively similar spectra in the lower panels.  The fully conjugated
ligand, however, pushes the peak in the lowest frequency band from
$\sim 29 \mathrm{cm}^{-1}$ to $\sim 55 \mathrm{cm}^{-1}$, yielding
significant overlap with the density of states in the nanoparticle.
This ligand also increases the overlap with the solvent density of
states in a band between 280 and 380 $\mathrm{cm}^{-1}$.  This
provides some physical basis for the high interfacial conductance
observed for the fully conjugated $C_8$ and $C_{12}$ ligands.

The similarity between the density of states for the alkanethiolate
and penultimate ligands also helps explain why the interfacial
conductance is nearly the same for these two ligands, particularly at
longer chain lengths.

\section{Discussion}

The chemical bond between the metal and the ligand introduces
vibrational overlap that is not present between the bare metal surface
and solvent. Thus, regardless of ligand identity or chain length, the
presence of a half-monolayer ligand coverage yields a higher
interfacial thermal conductance value than the bare nanoparticle.  The
mechanism for the varying conductance for the different ligands is
somewhat less clear.  Ligand-based alterations to vibrational density
of states is a major contributor, but some of the ligands can disrupt
the crystalline structure of the smaller nanospheres, while others can
re-order the interfacial solvent and alter the interpenetration
profile between ligand and solvent chains. Further work to separate
the effects of ligand-solvent interpenetration and surface
reconstruction is clearly needed for a complete picture of the heat
transport in these systems.

\begin{acknowledgments}
  Support for this project was provided by the National Science Foundation
  under grant CHE-1362211. Computational time was provided by the
  Center for Research Computing (CRC) at the University of Notre Dame.
\end{acknowledgments}

\newpage
%

\end{document}


\title{Supplemental Material for: Interfacial Thermal Conductance of Thiolate-Protected
  Gold Nanospheres}
\author{Kelsey M. Stocker}
\author{Suzanne M. Neidhart}
\author{J. Daniel Gezelter}
\email{gezelter@nd.edu}
\affiliation{Department of Chemistry and Biochemistry, University of
  Notre Dame, Notre Dame, IN 46556}
\date{\today}

\begin{abstract}
  This document supplies force field parameters for the united-atom
  sites, bond, bend, and torsion parameters, as well as the cross
  interactions between the united-atom sites and the gold atoms. These
  parameters were used in the simulations presented in the main text.
\end{abstract}

\maketitle

Gold -- gold interactions were described by the quantum Sutton-Chen
(QSC) model.\cite{Qi:1999ph} The hexane solvent is described by the
TraPPE united atom model,\cite{TraPPE-UA.alkanes} where sites are
located at the carbon centers for alkyl groups. Bonding interactions
were used for intra-molecular sites closer than 3 bonds. Effective
Lennard-Jones potentials were used for non-bonded interactions.

\begin{table}[h]
\bibpunct{}{}{,}{n}{}{,}
\centering
\caption{Non-bonded interaction parameters (including cross interactions with Au atoms). \label{tab:atypes}}
\begin{tabular}{ c|cccccl }
 \toprule
Site & mass & $\sigma_{ii}$ & $\epsilon_{ii}$ & $\sigma_{\ce{Au}-i}$ & $\epsilon_{\ce{Au}-i}$  & source \\
     & (amu)& (\AA)        & (kcal/mol)     & (\AA)             &  (kcal/mol)          &  \\
 \colrule
 \ce{CH3}    & 15.04    & 3.75  & 0.1947 & 3.54   & 0.2146 & Refs. \protect\cite{TraPPE-UA.alkanes}, \protect\cite{vlugt:cpc2007154} and \protect\cite{landman:1998}\\
 \ce{CH2}    & 14.03    & 3.95  & 0.09141& 3.54   & 0.1749 & Refs. \protect\cite{TraPPE-UA.alkanes}, \protect\cite{vlugt:cpc2007154} and \protect\cite{landman:1998}\\
 CHene       & 13.02    & 3.73  & 0.09340& 3.4625 & 0.1680 & Refs. \protect\cite{TraPPE-UA.alkylbenzenes}, \protect\cite{vlugt:cpc2007154} and \protect\cite{landman:1998}\\
 S           & 32.0655  & 4.45  & 0.2504 & 2.40   & 8.465  & Refs. \protect\cite{landman:1998} ($\sigma$) and \protect\cite{vlugt:cpc2007154} ($\epsilon$) \\
 CHar        & 13.02    & 3.695 & 0.1004 & 3.4625 & 0.1680 & Refs. \protect\cite{TraPPE-UA.alkylbenzenes} and \protect\cite{vlugt:cpc2007154}\\
 \ce{CH2ar}  & 14.03    & 3.695 & 0.1004 & 3.4625 & 0.1680 & Refs. \protect\cite{TraPPE-UA.alkylbenzenes} and \protect\cite{vlugt:cpc2007154}\\
 \botrule
\end{tabular}
\bibpunct{[}{]}{,}{n}{,}{,}
\end{table}

The TraPPE-UA force field includes parameters for thiol
molecules\cite{TraPPE-UA.thiols} which were used for the
alkanethiolate molecules in our simulations.  To derive suitable
parameters for butanethiolate adsorbed on Au(111) surfaces, we adopted
the S parameters from Luedtke and Landman\cite{landman:1998} and
modified the parameters for the CTS atom to maintain charge neutrality
in the molecule.

Bonds are typically rigid in TraPPE-UA, so although we used
equilibrium bond distances from TraPPE-UA, for flexible bonds, we
adapted bond stretching spring constants from the OPLS-AA force
field.\cite{Jorgensen:1996sf}

\begin{table}[h]
\bibpunct{}{}{,}{n}{}{,}
\centering
\caption{Bond parameters. \label{tab:bond}}
\begin{tabular}{ cc|ccl }
 \toprule
 $i$&$j$ & $r_0$ & $k_\mathrm{bond}$ & source \\
    &    & (\AA) & $(\mathrm{~kcal/mole/\AA}^2)$ & \\
 \colrule
\ce{CH3}   & \ce{CH3} &	1.540	& 536  & Refs. \protect\cite{TraPPE-UA.alkanes} and \protect\cite{Jorgensen:1996sf}\\
\ce{CH3}   & \ce{CH2} &	1.540	& 536  & Refs. \protect\cite{TraPPE-UA.alkanes} and \protect\cite{Jorgensen:1996sf} \\
\ce{CH2}   & \ce{CH2} &	1.540	& 536  & Refs. \protect\cite{TraPPE-UA.alkanes} and \protect\cite{Jorgensen:1996sf} \\
CHene      & CHene    & 1.330   & 1098 & Refs. \protect\cite{TraPPE-UA.alkylbenzenes} and \protect\cite{Jorgensen:1996sf}\\
\ce{CH3}   & CHene    & 1.540   & 634  & Refs. \protect\cite{TraPPE-UA.alkylbenzenes} and \protect\cite{Jorgensen:1996sf} \\
\ce{CH2}   & CHene    & 1.540   & 634  & Refs. \protect\cite{TraPPE-UA.alkylbenzenes} and \protect\cite{Jorgensen:1996sf} \\
S          & \ce{CH2} & 1.820   & 444  & Refs. \protect\cite{TraPPE-UA.thiols} and \protect\cite{Jorgensen:1996sf} \\
CHar       & CHar     & 1.40    & 938  & Refs. \protect\cite{TraPPE-UA.alkylbenzenes} and \protect\cite{Jorgensen:1996sf} \\
CHar       & \ce{CH2} & 1.540   & 536  & Refs. \protect\cite{TraPPE-UA.alkylbenzenes} and \protect\cite{Jorgensen:1996sf}\\
CHar       & \ce{CH3} & 1.540   & 536  & Refs. \protect\cite{TraPPE-UA.alkylbenzenes} and \protect\cite{Jorgensen:1996sf}\\
\ce{CH2ar} & CHar     & 1.40    & 938  & Refs. \protect\cite{TraPPE-UA.alkylbenzenes} and \protect\cite{Jorgensen:1996sf} \\
S          & CHar     &	1.80384	& 527.951 & This Work \\
 \botrule
\end{tabular}
\bibpunct{[}{]}{,}{n}{,}{,}
\end{table}

To describe the interactions between metal (Au) and non-metal atoms,
potential energy terms were adapted from an adsorption study of alkyl
thiols on gold surfaces by Vlugt, \textit{et
  al.}\cite{vlugt:cpc2007154} They fit an effective pair-wise
Lennard-Jones form of potential parameters for the interaction between
Au and pseudo-atoms CH$_x$ and S based on a well-established and
widely-used effective potential of Hautman and Klein for the Au(111)
surface.\cite{hautman:4994}

Parameters not found in the TraPPE-UA force field for the
intramolecular interactions of the conjugated and the penultimate
alkenethiolate ligands were calculated using constrained geometry
scans using the B3LYP functional~\cite{Becke:1993kq,Lee:1988qf} and
the 6-31G(d,p) basis set. Structures were scanned starting at the
minimum energy gas phase structure for small ($C_4$) ligands.  Only
one degree of freedom was constrained for any given scan -- all other
atoms were allowed to minimize subject to that constraint.  The
resulting potential energy surfaces were fit to a harmonic potential
for the bond stretching,
\begin{equation}
V_\mathrm{bond} = \frac{k_\mathrm{bond}}{2} \left( r - r_0 \right)^2,
\end{equation}
and angle bending potentials,
\begin{equation}
V_\mathrm{bend} = \frac{k_\mathrm{bend}}{2} \left(\theta - \theta_0\right)^2.
\end{equation}
Torsional potentials were fit to the TraPPE torsional function,
\begin{equation}
V_\mathrm{tor} = c_0 + c_1  \left(1 + \cos\phi \right) + c_2  \left(1 - \cos 2\phi \right) + c_3  \left(1 + \cos 3 \phi \right).
\end{equation}

For the penultimate thiolate ligands, the model molecule used was
2-Butene-1-thiol, for which one bend angle (\ce{S-CH2-CHene}) was
scanned to fit an equilibrium angle and force constant, as well as one
torsion (\ce{S-CH2-CHene-CHene}).  The parameters for these two
potentials also served as model for the longer conjugated thiolate
ligands which require bend angle parameters for (\ce{S-CH2-CHar}) and
torsion parameters for (\ce{S-CH2-CHar-CHar}).

For the $C_4$ conjugated thiolate ligands, the model molecule for the
quantum mechanical calculations was 1,3-Butadiene-1-thiol.  This
ligand required fitting one bond (\ce{S-CHar}), and one bend angle
(\ce{S-CHar-CHar}).

The geometries of the model molecules were optimized prior to
performing the constrained angle scans, and the fit values for the
bond, bend, and torsional parameters were in relatively good agreement
with similar parameters already present in TraPPE.

\begin{table}[h]
\bibpunct{}{}{,}{n}{,}{,}
\centering
\caption{Bend angle parameters. The central atom in the bend is atom $j$.\label{tab:bend}}
\begin{tabular}{ ccc|ccl }
\toprule
 $i$&$j$&$k$ & $\theta_0$ & $k_\mathrm{bend}$ & source\\
    &   &    & ($\degree$) & (kcal/mol/rad\textsuperscript{2}) & \\
 \colrule
\ce{CH2} & \ce{CH2} & S         & 114.0   &   124.20& Ref. \protect\cite{TraPPE-UA.thiols}\\ 
\ce{CH3} & \ce{CH2} & \ce{CH2}  & 114.0   &   124.20& Ref. \protect\cite{TraPPE-UA.thiols}\\ 
\ce{CH2} & \ce{CH2} & \ce{CH2}  & 114.0   &   124.20& Ref. \protect\cite{TraPPE-UA.thiols}\\ 
CHene    & CHene    & \ce{CH3}  & 119.7   &   139.94& Ref. \protect\cite{TraPPE-UA.alkylbenzenes}\\ 
CHene    & CHene    & \ce{CH2}  & 119.7   &   139.94& Ref. \protect\cite{TraPPE-UA.alkylbenzenes}\\ 
\ce{CH2} & \ce{CH2} & CHene     & 114.0   &   124.20& Ref. \protect\cite{TraPPE-UA.alkylbenzenes}\\ 
CHar     & CHar     & CHar      & 120.0   &   126.0 & Refs. \protect\cite{TraPPE-UA.alkylbenzenes} and \protect\cite{Jorgensen:1996sf}\\ 
CHar     & CHar     & \ce{CH2}  & 120.0   &   140.0 & Refs. \protect\cite{TraPPE-UA.alkylbenzenes} and \protect\cite{Jorgensen:1996sf}\\
CHar     & CHar     & \ce{CH3}  & 120.0   &   140.0 & Refs. \protect\cite{TraPPE-UA.alkylbenzenes} and \protect\cite{Jorgensen:1996sf}\\
CHar     & CHar     & \ce{CH2ar}& 120.0   &   126.0 & Refs. \protect\cite{TraPPE-UA.alkylbenzenes} and \protect\cite{Jorgensen:1996sf}\\
S        & \ce{CH2} & CHene     & 109.97  &  127.37 & This work  \\
S        & \ce{CH2} & CHar      & 109.97  &  127.37 & This work  \\
S        & CHar     & CHar      & 123.911 & 138.093 & This work  \\
 \botrule
\end{tabular}
\bibpunct{[}{]}{,}{n}{,}{,}
\end{table}

\begin{table}[h]
\bibpunct{}{}{,}{n}{,}{,}
\centering
\caption{Torsion parameters. The central atoms for each torsion are atoms $j$ and $k$,
  and wildcard atom types are denoted by ``X''.  All $c_n$ parameters
  have units of kcal/mol. The torsions around carbon-carbon double bonds
  are harmonic and assume a trans (180$\degree$) geometry.  The force
  constant for this torsion is given in $\mathrm{kcal~mol~}^{-1}\mathrm{degrees}^{-2}$.  \label{tab:torsion}}
\begin{tabular}{ cccc|ccccl }
\toprule
 $i$&$j$&$k$&$l$& $c_0$&$c_1$& $c_2$ & $c_3$ & source\\
 \colrule
\ce{CH3} & \ce{CH2} & \ce{CH2} & \ce{CH2} & 0.0     & 0.7055   & -0.13551 &  1.5725    & Ref. \protect\cite{TraPPE-UA.alkanes}\\
\ce{CH2} & \ce{CH2} & \ce{CH2} & \ce{CH2} & 0.0     & 0.7055   & -0.13551 &  1.5725    & Ref. \protect\cite{TraPPE-UA.alkanes}\\
\ce{CH2} & \ce{CH2} & \ce{CH2} & S        & 0.0     & 0.7055   & -0.13551 &  1.5725    & Ref. \protect\cite{TraPPE-UA.thiols}\\ \colrule
X        & CHene    & CHene    & X        & \multicolumn{4}{c}{\multirow{2}{*}{$V = \frac{0.008112}{2} (\phi - 180.0)^2$}} & \multirow{2}{*}{Ref. \protect\cite{TraPPE-UA.alkylbenzenes}} \\
X        & CHar     & CHar     & X        &         & & & & \\ \colrule
\ce{CH2} & \ce{CH2} & CHene    & CHene    & 1.368   & 0.1716   & -0.2181  &  -0.56081  & Ref. \protect\cite{TraPPE-UA.alkylbenzenes}\\
\ce{CH2} & \ce{CH2} & \ce{CH2} & CHene    & 0.0     & 0.7055   & -0.13551 &   1.5725   & Ref. \protect\cite{TraPPE-UA.alkylbenzenes}\\
CHene    & CHene    & \ce{CH2} & S        & 3.20753 & 0.207417 & -0.912929&  -0.958538 & This work \\
CHar     & CHar     & \ce{CH2} & S        & 3.20753 & 0.207417 & -0.912929&  -0.958538 & This work \\
 \botrule
\end{tabular}
\bibpunct{[}{]}{,}{n}{,}{,}
\end{table}

\newpage
%